\documentclass[prr,aps,twocolumn,amsmath,amsmath,amssymb,
    superscriptaddress,longbibliography]{revtex4-1}

\usepackage[T1]{fontenc}
\usepackage[utf8]{inputenc}
\usepackage{graphicx}
\usepackage[dvipsnames]{xcolor}
\usepackage[colorlinks,linkcolor=blue,citecolor=blue,urlcolor=blue]{hyperref}
\usepackage{stix}
\DeclareUnicodeCharacter{2009}{\,}

\makeatletter
\usepackage[english]{babel}
\usepackage{enumitem}
\usepackage{bm}
\usepackage{braket}

\newcommand{\BiSe}{\mathrm{Bi}_2\mathrm{Se}_3}

\renewcommand{\vec}[1]{\mathbf{#1}}
\newcommand{\gvec}[1]{\boldsymbol{#1}}

\makeatother

\usepackage{babel}
\begin{document}
\title{\selectlanguage{english}%
Reexamining Circular Dichroism in Photoemission From a Topological Insulator}

\author{Ittai Sidilkover}
\affiliation{School of Physics and Astronomy, Faculty of Exact Sciences, Tel Aviv University, Tel‐Aviv, 6997801, Israel}
\affiliation{Center for Light‐Matter Interaction, Tel Aviv University, Tel‐Aviv, 6997801, Israel}
\author{Yun Yen}
\affiliation{PSI Center for Scientific Computing, Theory and Data, 5232 Villigen PSI, Switzerland}
\affiliation{École Polytechnique Fédérale de Lausanne (EPFL), Switzerland}
\author{Sunil Wilfred D’Souza} 
\affiliation{New Technologies Research Centre, University of West Bohemia, Univerzitní 8, CZ-306 14 Pilsen, Czech Republic}
\author{Jakub Schusser} 
\affiliation{New Technologies Research Centre, University of West Bohemia, Univerzitní 8, CZ-306 14 Pilsen, Czech Republic}
\author{Aki Pulkkinen} 
\affiliation{New Technologies Research Centre, University of West Bohemia, Univerzitní 8, CZ-306 14 Pilsen, Czech Republic}
\author{Costel R. Rotundu}
\affiliation{Stanford Institute for Materials and Energy Sciences, SLAC National Accelerator Laboratory, 2575 Sand Hill Road, Menlo Park, California 94025, USA}
\affiliation{Geballe Laboratory for Advanced Materials, Stanford University, Stanford, California 94305, USA}
\author{Makoto Hashimoto}
\affiliation{Stanford Synchrotron Radiation Lightsource, SLAC National Accelerator Laboratory, Menlo Park, California 94025, USA}
\author{Donghui Liu}
\affiliation{Stanford Synchrotron Radiation Lightsource, SLAC National Accelerator Laboratory, Menlo Park, California 94025, USA}
\author{Zhi-Xun Shen}
\affiliation{Stanford Institute for Materials and Energy Sciences, SLAC National Accelerator Laboratory, 2575 Sand Hill Road, Menlo Park, California 94025, USA}
\affiliation{Geballe Laboratory for Advanced Materials, Stanford University, Stanford, California 94305, USA}
\affiliation{Department of Physics, Stanford University, Stanford, California 94305, USA}
\author{J\'{a}n Min\'{a}r}
\affiliation{New Technologies Research Centre, University of West Bohemia, Univerzitní 8, CZ-306 14 Pilsen, Czech Republic}
\author{Michael Schüler}
\email{michael.schueler@psi.ch}
\affiliation{PSI Center for Scientific Computing, Theory and Data, 5232 Villigen PSI, Switzerland}
\affiliation{University of Fribourg, Department of Physics, University of Fribourg, CH-1700 Fribourg, Switzerland}
\author{Hadas Soifer}
\email{hadassoifer@tauex.tau.ac.il}
\affiliation{School of Physics and Astronomy, Faculty of Exact Sciences, Tel Aviv University, Tel‐Aviv, 6997801, Israel}
\affiliation{Center for Light‐Matter Interaction, Tel Aviv University, Tel‐Aviv, 6997801, Israel}
\author{Jonathan A. Sobota}
\email{sobota@slac.stanford.edu}
\affiliation{Stanford Institute for Materials and Energy Sciences, SLAC National Accelerator Laboratory, 2575 Sand Hill Road, Menlo Park, California 94025, USA}

\selectlanguage{english}%
\begin{abstract}
The orbital angular momentum (OAM) of electron states is an essential ingredient for topological and quantum geometric quantities in solids. For example, Dirac surface states with helical spin- and orbital-angular momenta are a hallmark of a 3D topological insulator. Angle-resolved photoemission spectroscopy (ARPES) with variable circular light polarization, known as circular dichroism (CD), has been assumed to be a direct probe of OAM and, by proxy, of the Berry curvature of electronic bands in energy- and momentum-space. Indeed, topological surface states have been shown to exhibit angle-dependent CD (CDAD), and more broadly, CD is often interpreted as evidence of spin-orbit coupling. Meanwhile, it is well-established that CD originates from the photoemission matrix elements, which can have extrinsic contributions related to the experimental geometry and the inherently broken inversion symmetry at the sample surface. Therefore, it is important to broadly examine CD-ARPES to determine the scenarios in which it provides a robust probe of intrinsic material physics. We performed CD-ARPES on the canonical topological insulator $\BiSe$ over a wide range of incident photon energies. Not only do we observe angle-dependent CD in the surface states, as expected, but we also find CD of a similar magnitude in virtually all bulk bands. Since OAM is forbidden by inversion symmetry in the bulk, we conclude this originates from symmetry-breaking in the photoemission process. Comparison with theoretical calculations supports this view and suggests that “hidden” OAM -- localized to atomic sites within each unit cell -- contributes significantly. Additional effects, including inter-atomic interference and final-state resonances, are responsible for the rapid variation of the CDAD signal with photon energy.

\end{abstract}
\maketitle

\section{Introduction}

Electronic states in crystalline solids are described by energy bands $\varepsilon(\vec{k})$ and Bloch wavefunctions $| u_{\vec{k}} \rangle$ defined over reciprocal space $\vec{k}$. 
The notions of Berry connection and curvature, which describe the phase evolution of the Bloch wavefunction in momentum space, have led to a paradigm shift in condensed matter physics. Their far-reaching consequences include the quantum Hall effect \cite{zhang_experimental_2005}, its nonlinear variant \cite{sodemann_quantum_2015}, 
light-matter coupling~\cite{ahn_riemannian_2022},
and the identification of classes of topological materials \cite{yan_topological_2012} such as Weyl semimetals. The Berry connection is also a key quantity in the formulation of the modern theory of polarization and magnetization \cite{souza_dichroic_2008,xiao_berry_2010,resta_electrical_2010}. The Berry curvature describes the intrinsic properties of the band originating in the self-rotation of an electronic wavepacket, such as the orbital angular momentum (OAM), which shares the same symmetries \cite{ma_chiral_2015}. OAM allows for non-vanishing orbital magnetization even in the absence of magnetic atoms, as long as time reversal or inversion symmetries are broken by other means \cite{takahashi_berry_2015,resta_magnetic_2020}. 
On the other hand, each atom can host non-vanishing local atomic OAM~\cite{ryoo_hidden_2017}, which may contribute to the global OAM depending on symmetry considerations. Both types of OAM are closely related to the underlying topology, with recent studies aiming to extract information about the Berry curvature and the quantum geometric tensors from it \cite{yen_controllable_2024, brinkman_chirality-driven_2024,kang_measurements_2025}. Additionally, the observation of the orbital Hall effect (OHE) \cite{choi_observation_2023}, and the field of orbitronics in general~\cite{go_intrinsic_2018-1, choi_observation_2023, rappoport_first_2023}, has sparked a new interest in OAM and its dynamics in solids. Being of such importance, OAM is a coveted quantity to be probed experimentally. 

\begin{figure}
\centering \includegraphics[width=0.5\textwidth]{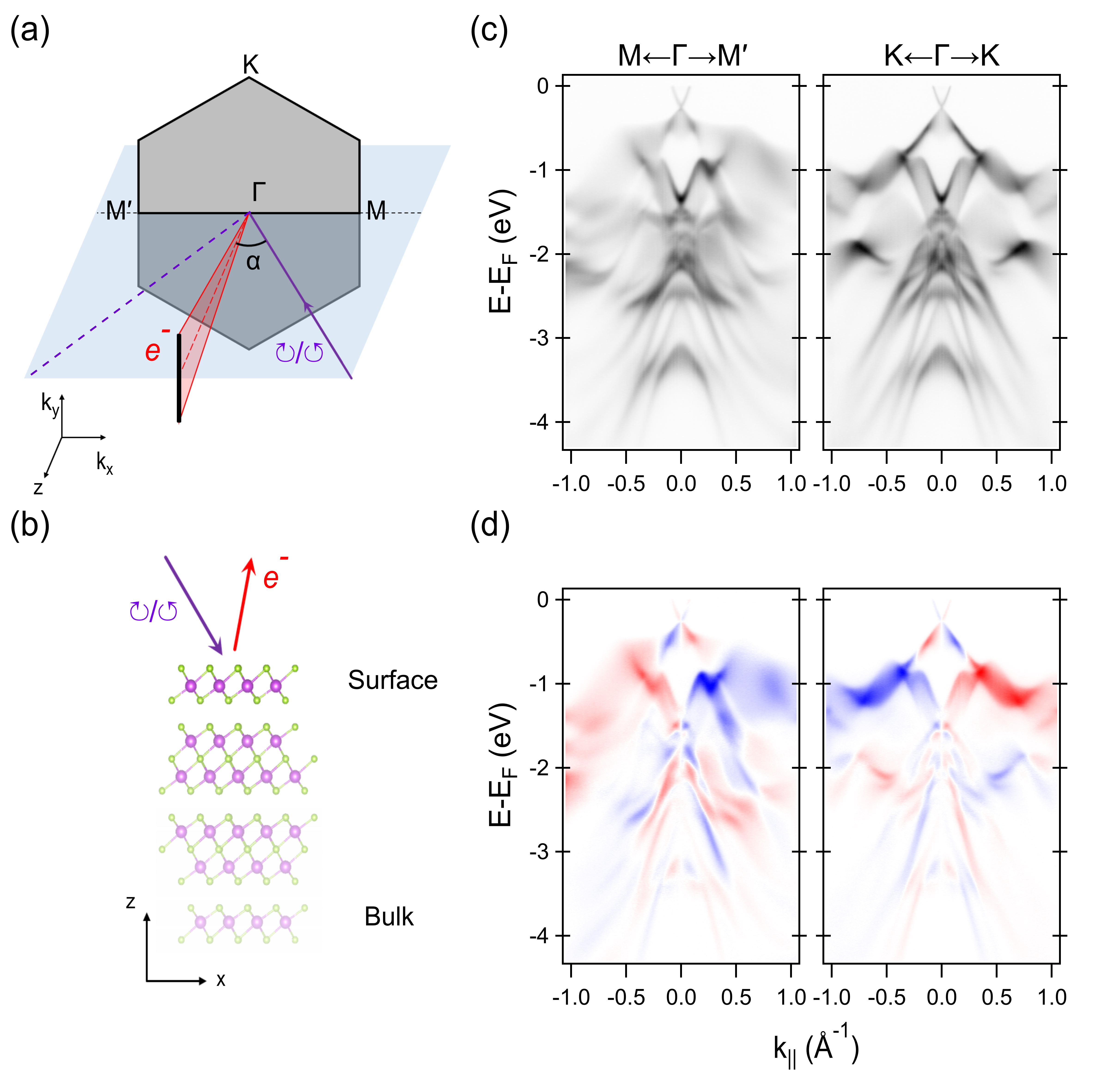}
\caption{(a) Schematic drawing of the experimental geometry of the ARPES experiment. The angle $\alpha$ indicates the incidence direction of the circularly polarized light (purple line) compared to the normal emission of the electrons (red region) towards the slit of the analyzer (black line). The relevant mirror plane of the crystal is shown in blue. (b) Side view of the surface crystal structure of $\BiSe$. (c) Sum ($I^{(+)}+I^{(-)}$) and (d) difference ($I^{(+)}-I^{(-)}$)  of ARPES spectra obtained with circularly polarized light, along the $M-\Gamma-M'$ (left) and $K-\Gamma-K$ (right) directions. The spectra were obtained with $45$~eV circularly polarized light.}
\label{fig:Exp_Geo}
\end{figure}

Angle-resolved photoemission spectroscopy (ARPES) has proven to be a central experimental technique in studying quantum and topological materials with its ability to probe electronic band structure directly \cite{gedik_photoemission_2017,lv_angle-resolved_2019,sobota_angle-resolved_2021}. Observation of surface Dirac cones in topological insulators \cite{xia_observation_2009,chen_experimental_2009, sobota_direct_2013} or Weyl points and Fermi-arcs in Weyl semimetals \cite{xu_discovery_2015,yang_weyl_2015, liu_magnetic_2019, rossi_electronic_2021} are regarded as clear indicators for the topological nature of a compound. However, being a phase-related quantity, the Berry curvature itself is inaccessible via conventional ARPES, a spectroscopic method in which only the square of the matrix element is captured \cite{schuler_polarization-modulated_2022}. 

A connection between the Berry curvature and OAM of materials to their photoemission response under circularly polarized light has been established recently \cite{de_juan_quantized_2016,schuler_local_2020,pozo_quantization_2019}. In particular, circular dichroism in the angular distribution (CDAD), defined as the difference between two ARPES spectra obtained with right- and left-circularly polarized photons, has emerged as a highly informative experimental observable~\cite{figgemeier_tomographic_2025}. In a simplified photoemission picture and for specific two-dimensional materials, CDAD is proportional to the Berry curvature or OAM of the band from which the electrons were emitted \cite{fedchenko_4d_2019,park_orbital-angular-momentum_2011,schuler_local_2020, cho_studying_2021,wang_observation_2011}. In practice, however, extracting OAM information is generally difficult due to several other mechanisms that can give rise to CDAD and obstruct the underlying OAM pattern.


Recent theoretical and experimental works \cite{razzoli_selective_2017,beaulieu_revealing_2020,yen_controllable_2024,brinkman_chirality-driven_2024,Schusser2024} demonstrate that ARPES under certain conditions can reveal \textit{hidden} OAM and spin components \cite{zhang_hidden_2014,riley_direct_2014,razzoli_selective_2017,tu_direct_2020}. This is because the finite inelastic mean-free path of the photoelectron (IMFP) leads to an exponential attenuation factor of the signal, even on the scale of a unit cell, which breaks inversion symmetry and reveals the local atomic OAM. 
The consequences of atomic OAM were recently explored in ${\mathrm{W}\mathrm{Se}_2}$ under different photon energy regimes, probing CDAD from mostly the topmost layer \cite{beaulieu_revealing_2020,cho_experimental_2018}, or from bulk layers \cite{Schusser2024}; and in bulk chiral semimetals \cite{yen_controllable_2024,brinkman_chirality-driven_2024}, where it accompanies the signal originating from the magnetic-chiral structure of the bulk crystal.

In this work, we present CDAD measurements of the bulk inversion-symmetric topological insulator $\BiSe$ exhibiting both bulk and surface states. Whereas CDAD of the topological surface states has been used in the past as evidence of the OAM or spin texture of the Dirac cone \cite{park_chiral_2012,vidal_photon_2013}, here we show that the bulk states exhibit CDAD of the same magnitude. Since the bulk of $\BiSe$ is inversion (and time-reversal) symmetric, precluding global OAM, this indicates a substantial contribution of hidden OAM. 
We explore the different mechanisms responsible for the experimental dichroic signal of bulk and surface states and determine their significance and strength by comparing model calculations and first-principle simulations with CD-ARPES experiments over a wide range of photon energies.

\section{Experimental results}
\label{s:exp}
ARPES measurements were performed at beam line 5-2 of SSRL at SLAC equipped with a Scienta D80 hemispherical electron analyzer. Samples of pristine $\BiSe$ were oriented to the $\Gamma-K$ and $\Gamma-M$ directions and cleaved in-situ at $T=10$~K and base pressure $<5\cdot 10^{-11}$~Torr. ARPES spectra were obtained with right- and left-handed polarized XUV photons within the $25.5-50$~eV energy range. The light incidence is at $\alpha=50^\circ$ as shown in Fig.\ref{fig:Exp_Geo}a, impinging the crystal surface as shown in Fig.\ref{fig:Exp_Geo}b. The photon flux for each polarization and photon energy was recorded independently and was used to normalize each spectrum.

\begin{figure*}
\includegraphics[width=1\textwidth]{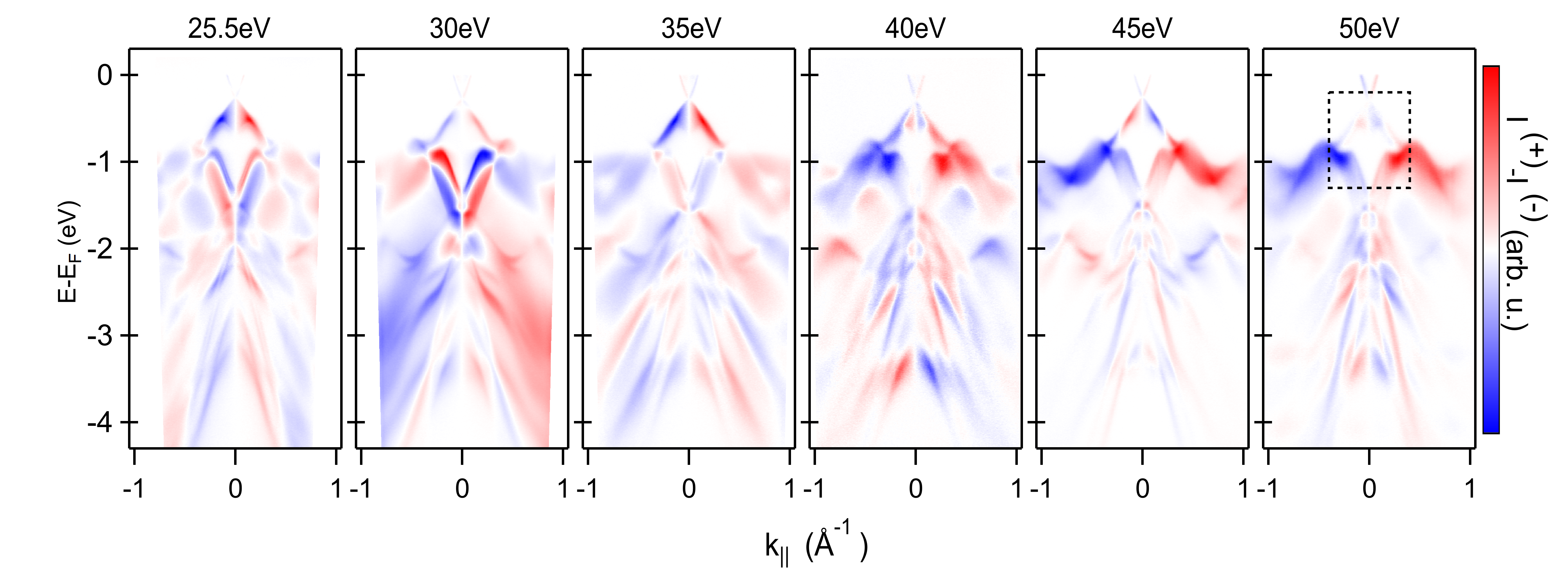}
\caption{CDAD ($I^{(+)}-I^{(-)}$) along the $K-\Gamma-K$ direction of $\BiSe$, obtained with circularly polarized light at $25.5-50$~eV photon energies. The dashed region in the right panel marks the zoomed-in region discussed in Fig.\ref{fig:intra_inter} for all photon energies.}
\label{fig:wide_CDAD}
\end{figure*}

The polarization-summed ARPES spectra ($I^{(+)}+I^{(-)}$, with $I^{(\pm)}$ noting the ARPES intensity for right/left circularly polarized light) along the two high-symmetry axes of the Brillouin zone are shown in Fig.\ref{fig:Exp_Geo}c and the CDAD ($I^{(+)}-I^{(-)}$) is plotted in Fig.\ref{fig:Exp_Geo}d. As expected, the topological Dirac surface state, located near the $\Gamma$-point within 0.5~eV of the Fermi level, exhibits a dichroic signal which switches sign at the Dirac point ($\simeq-0.3$~eV). Surprisingly, the bulk bands also exhibit a strong circular dichroism that naively should vanish as both inversion and time-reversal symmetries are present in the bulk of the sample. Nonetheless, it has a comparable magnitude as the CD of the topological Dirac surface states. The CDAD pattern respects some of the crystal symmetries, with a clear antisymmetric structure along $\Gamma-K$ due to the mirror plane along $\Gamma-M$ (the blue horizontal plane in Fig.\ref{fig:Exp_Geo}a), whereas along the non-symmetric $M-\Gamma-M'$ this anti-symmetry is broken. 

Furthermore, the CDAD pattern varies drastically as the photon energy is changed between 25.5 to 50 eV, as shown in Fig.\ref{fig:wide_CDAD}. The dichroic signal is maximal for bands near $\Gamma$ at low photon energies, while for higher ones, the greater contrast is at higher momentum regions. Several sign flips of the dichroic signal can also be seen as the photon energy increases, especially around $\Gamma$. 
These counter-intuitive results suggest additional mechanisms that contribute to the dichroism in the photoemission process, rooted in hidden OAM. 

\section{Theoretical and numerical methods to simulate CDAD in ARPES}

Describing photoemission accurately is challenging. For semi-infinite systems, the Kohn-Korringa-Rostoker (KKR) Green's function approach provides the most efficient first-principle framework for simulating ARPES~\cite{ebert_calculating_2011}. The method naturally incorporates multiple scattering effects into the final state of photoemission, the time-reversed LEED state. It is, however, not straightforward to establish a direct connection between features in the ARPES signals and the orbital texture in this method. The Wannier-ARPES method~\cite{beaulieu_unveiling_2021-1,yen_first-principle_2024}, on the other hand, models the photoemission process as a coherent sum of partial waves emitted from each atom~\cite{moser_toy_2023}. While the final state is only approximate in this method, a direct orbital-resolved analysis is straightforward. For the qualitative understanding of the CDAD observed in the experiments, we thus use the Wannier-ARPES approach, while subtle effects in the CDAD from the surface states are analyzed within the KKR framework. In the following we use atomic units (a.u.) unless stated otherwise.


\subsection{Photoemission simulation: Wannier-ARPES approach}
\label{s:wannier_arpes}

The ARPES intensity can be computed using Fermi's Golden rule as
    \begin{align}
        \label{eq:fermi_golden}
        I^{(\pm)}(\vec{k}_\parallel, E) = \sum_\alpha  |M^{(\pm)}_\alpha(\vec{k}_\parallel,E)|^2 \delta(\varepsilon_\alpha(\vec{k}) + \hbar\omega - E - \Phi) \ , 
    \end{align}
    where, as before, $I^{(\pm)}$ describes the ARPES intensity with respect to right/left circularly polarized light at kinetic energy $E$ and in-plane crystal momentum $\vec{k}_\parallel$. $\Phi$ denotes the work function. $M_\alpha(\vec{k}_\parallel,E)$ is the light-matter coupling matrix element connecting the photoelectron final state $\ket{\chi_{\vec{p}}}$ and Bloch state $|\psi_{\vec{k}\alpha} \rangle$, which can be expressed in the dipole gauge as 
    \begin{align}
        \label{eq:lm_me}
        M_{\alpha}(\vec{k}_{\parallel},E) &= \gvec{\epsilon} \cdot \langle \chi_{\vec{p}} | \hat{\mathbf{r}} | \psi_{\vec{k}\alpha} \rangle \ ,
    \end{align}
    where $\gvec{\epsilon}$ is the light polarization vector. The photoelectron momentum $\vec{p}=(\vec{k}_{\parallel},p_z)$ is fixed by the energy conservation $p_z^2/2 + k_\parallel^2/2 = E \ $. We compute the matrix elements within the Wannier-ARPES approach~\cite{yen_first-principle_2024,beaulieu_revealing_2020,schuler_polarization-modulated_2022,day_computational_2019} (see Appendix~\ref{appendix:relation}), where the Bloch states are projected to local Wannier orbitals, and the matrix element within the atomic center approximation~\cite{yen_first-principle_2024} can be expanded as
    \begin{align}
        \label{eq:mel_dip_aca}
        M_\alpha(\vec{k}_{\parallel}, E) = \sqrt{N} \sum_j C_{j\alpha}(\vec{k}_{\parallel}) e^{-i\vec{p}\cdot \vec{r}_j} e^{z_j/\lambda} M^{\mathrm{orb}}_j(\vec{k}_{\parallel},E) \ ,
    \end{align}

    \begin{align}
        \label{eq:mel_dip_aca_orb}            M^{\mathrm{orb}}_j(\vec{k}_{\parallel},E) =  \int d\vec{r}\, \chi_{\vec{p}}^{*}(\vec{r}) 
            \vec{\epsilon}\cdot \vec{r}
            \phi_j(\vec{r})  \ . 
    \end{align}
    Here $C_{j\alpha}(\vec{k})$ is the projection of Bloch state $\alpha$ onto Wannier orbital $\phi_j(\vec{r})$. $N$ is the number of lattice sites. The effective photoelectron IMFP $\lambda$ is included as an attenuation factor, as shown in Fig.~\ref{fig:Local_OAM}(a). In principle, the IMFP varies with the electron kinetic energy and roughly follows the so-called universal curve \cite{seah_quantitative_1979}. In the photon energy range we focus on, the IMFP is on the order of one quintuple layer ($10\mathring{A}$) \cite{zhang_quintuple-layer_2009}. The details of computing the orbital matrix element with respect to the Wannier orbitals $\phi_j(\vec{r}) $ in Eq.~\eqref{eq:mel_dip_aca_orb} are discussed in Appendix \ref{appendix:relation}.

\subsection{How is OAM reflected in CDAD within the Wannier-ARPES approach?}

The \textit{global} OAM of an occupied Bloch state $\alpha$ can be computed with the modern theory of orbital magnetization \cite{souza_dichroic_2008,xiao_berry_2010,resta_magnetic_2020} as
    \begin{equation}
        \label{eq:modern_oam2}
        \mathbf{L}_{\alpha}(\mathbf{k}) = -i \sum_{\alpha' \in \mathrm{unocc}} (\varepsilon_{\alpha'}(\mathbf{k}) - \varepsilon_{\alpha}(\mathbf{k})) \vec{A}_{\alpha \alpha'}(\mathbf{k}) \times \vec{A}_{\alpha' \alpha}(\mathbf{k}) \ .
    \end{equation}
Here, $\vec{A}_{\alpha \alpha'}(\mathbf{k})$ denotes the Berry connection, which can be easily computed within the Wannier representation of the Bloch state~\cite{yates_spectral_2007}. In fact, Eq.\eqref{eq:modern_oam2} indicates the close relation between OAM and Berry curvature. For a two band crossing, both quantities are proportional to each other in the vicinity of the crossing node~\cite{schuler_local_2020, xiao_berry_2010}. 

On the other hand, \textit{local} OAM is defined by the expectation value of the OAM operator with respect to local atomic states, and can be formulated within the Wannier representation as  
    \begin{equation}
        \label{eq:local_oam}
        \mathbf{L}_{s,\alpha}(\vec{k}) = \sum_{m',m}  C_{(sm)\alpha}^{*}(\vec{k})  \mathbf{L}_{mm'} C_{(sm')\alpha}(\vec{k}) \ ,
    \end{equation}
where $C_{j\alpha}(\mathbf{k})=C_{(sm)\alpha}(\mathbf{k})$ is the projection of Bloch state $\alpha$ onto magnetic orbital $m$ at atomic site $s$. $\mathbf{L}_{mm'}=\langle m|(\hat{L}^{x},\hat{L}^{y},\hat{L}^{z})|m'\rangle$ denotes the atomic OAM with  matrix elements for OAM operator $\hat{L}^{\mu}$ along direction $\mu=x,y,z$, with respect to magnetic orbitals of magnetic quantum numbers $m$ and $m'$. 

In materials -- like Bi$_2$Se$_3$ -- where both time-reversal and inversion symmetries are observed, Kramer's degeneracy is enforced throughout the Brillouin zone, so all the bulk bands are doubly degenerate. As a result, the Berry curvature and \textit{global} OAM cancel when summing over each degenerate pair. Nevertheless, there can be non-zero \textit{local} OAM on each local atomic site 
which could give rise to CDAD in the presence of finite IMFP. To explain the microscopic mechanism, we first simplify the CDAD within Wannier-ARPES formulation to
    \begin{align}
        \label{eq:CDAD_theory_intensity}
        I_{\mathrm{CD}}(\vec{k}_\parallel&,E) = I^{(+)}(\vec{k}_\parallel, E) - I^{(-)}(\vec{k}_\parallel, E)\nonumber \\
        &\propto \sum_{sms'm'} C^*_{(sm)\alpha}(\vec{k}_\parallel)T_{sms'm'}(\vec{k}_\parallel, E)C_{(s'm')\alpha}(\vec{k}_\parallel) \ . 
    \end{align}
    Here $T_{sms'm'}(\vec{k}_\parallel, E)$, which includes all the matrix element details, is defined as
    \begin{align}
        \label{eq:T_tensor}
        T_{sms'm'}(&\vec{k}_\parallel, E) = e^{-i\vec{p}\cdot(\vec{r}_s -\vec{r}_{s'})} e^{(z_s+z_{s'})/\lambda}\nonumber \\ & \times [M_{sm}^{orb(+)*}M_{s'm'}^{orb(+)}-M_{sm}^{orb(-)*}M_{s'm'}^{orb(-)}] \ ,
    \end{align}
    where $\vec{r}_s$ denotes the position for atom $s$. The formulation allows us to decompose CD into two parts as

    \begin{align}
        \label{eq:CDAD_theory_intensity}
        I_{\mathrm{CD}}(\vec{k}_\parallel,E) 
        &=\sum_s I_{\mathrm{intra}}^{s}(\vec{k}_\parallel,E) + \sum_{ss'} I_{\mathrm{inter}}^{s \neq s'} (\vec{k}_\parallel,E) \ . 
    \end{align}
  The intra-atomic contribution $I_{\mathrm{intra}}^{s}$ is related to the local OAM of each of the atomic sites, and the inter-atomic contribution $I_{\mathrm{inter}}^{s \neq s'}$ includes interference from different atomic sites, adding a strong modulation to the signal with the phase factor $e^{-i\vec{p}\cdot(\vec{r}_s - \vec{r}_{s'})}$ in Eq.~\eqref{eq:T_tensor}~\cite{heider_geometry-induced_2023}. 


\section{Analyzing contributions to experimental CDAD}

\subsection{Observing CDAD in bulk bands: finite IMFP providing sensitivity to local OAM}  
\label{s:finite_cd}
\begin{figure}
    \centering
    \includegraphics[width=0.5\textwidth]{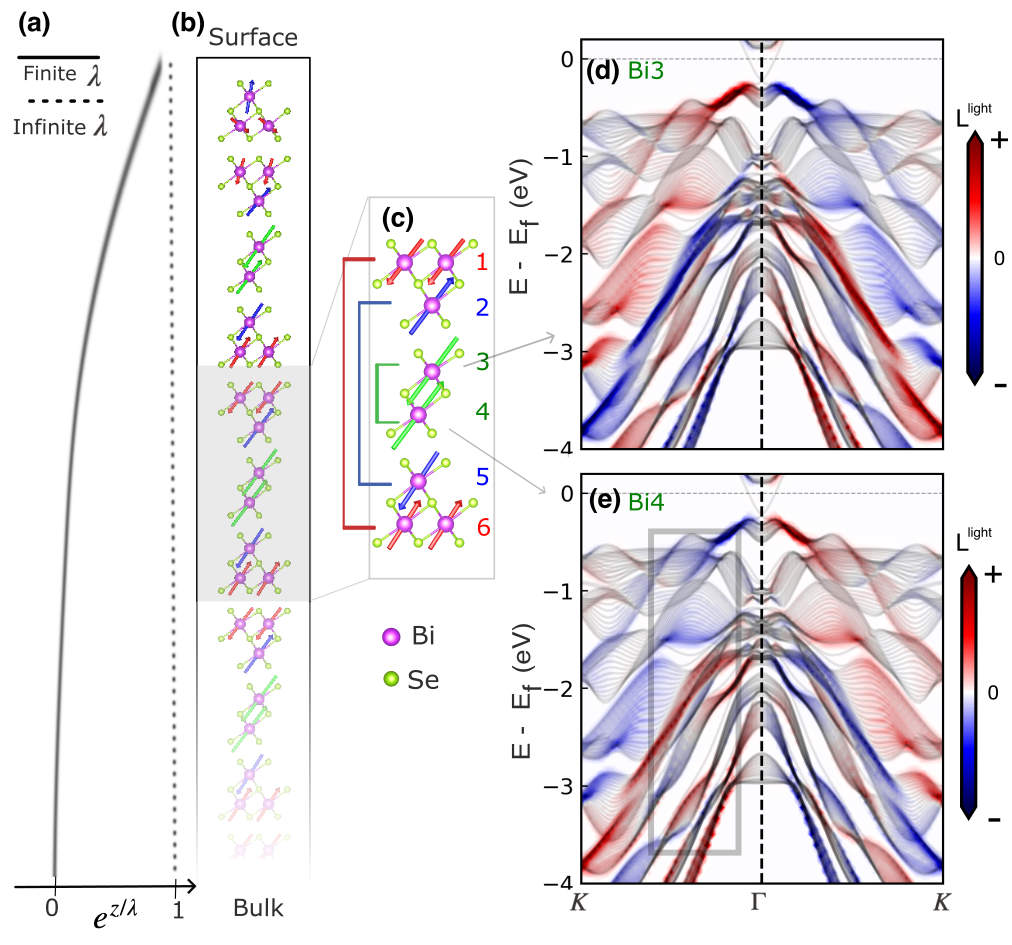}
    \caption{Local OAM and inversion symmetry. (a) Schematic of exponential profile with inelastic mean free path (IMFP) $\lambda$. (b) Calculated local OAM vectors for Bismuth atoms in a $\BiSe$ slab and (c) the zoomed-in image demonstrating inversion-symmetric OAM pairs of Bismuth atoms. Slab band resolved local OAM for (d) Bi3 and (e) Bi4 along the in-coming light direction in the zoomed-in cell. Note that the vectors in (b)-(c) are integrated over the range of the highlighted box in (e). }
    \label{fig:Local_OAM}
\end{figure}

We can now use this approach to explain the most prominent feature of the experimental data in Figs.~\ref{fig:Exp_Geo} and \ref{fig:wide_CDAD}: the abundance of strong CDAD in bulk bands of $\BiSe$, where global OAM is forbidden due to inversion symmetry. 
 Indeed, the local OAM of the atoms is finite \cite{ryoo_hidden_2017} (except for the Selenium at the inversion center), but with opposite orientation for inversion symmetric pairs, thus adding up to net zero global OAM. To visualize this, we performed an \textit{ab-initio} slab calculation and computed the local OAM vector for each Bismuth atom, as shown in Fig.~\ref{fig:Local_OAM}(b). The inversion symmetric pairs can be clearly observed in the zoomed image in Fig.~\ref{fig:Local_OAM}(c). Note that time-reversal symmetry is reflected in $\vec{k}$-space, leading to opposite local OAM for opposite $\vec{k}$, as shown in Fig.~\ref{fig:Local_OAM}(d) and (e) for the atoms labeled Bi3 and Bi4. It is clear that summing up the local OAM of these two atoms leads to net zero OAM at all $\vec{k}$. 
 


However, these considerations do not describe the signal in an actual ARPES experiment, where the presence of a surface and a finite IMFP imply that local quantities may not cancel. 
First, inversion symmetry is inherently broken at the surface, which can induce surface OAM~\cite{wawrzik_surface_2023}.
However, our calculations show that the impact of this effect on the CDAD of bulk states is negligible (see Appendix~\ref{appendix:layer_resolved}). Second, the photoemission exponential profile due to the finite IMFP (Fig.~\ref{fig:Local_OAM}a) leads to the breakdown of pair-wise cancellation of the atomic contributions, thus revealing the local atomic OAM in the CDAD. In the case of the inversion-symmetric Bismuth atom pair described above, photoemission from Bi4 is approximately suppressed by a factor of 2 compared to Bi3. Therefore the two contributions calculated in Fig.~\ref{fig:Local_OAM}(d) and (e) will not cancel each other. We conclude this is the principal origin for CDAD from bulk inversion-symmetric states. 

\subsection{Mechanisms underlying photon energy dependence of the CDAD}
\label{s:mechanism}

 The next striking experimental observation is the pronounced photon-energy dependence of the CDAD, as can be seen in Fig.\ref{fig:wide_CDAD}. The local OAM is a ground-state property of the material and does not depend on photon energy, and therefore cannot, on its own, explain this feature. In this section, we examine in detail different contributions to CDAD in $\BiSe$, and their relative significance for photon-energy dependence of the signal. 

First, we must consider that the probed $k_z$ in ARPES depends on photon energy, and the finite IMFP gives a finite $k_z$ resolution, as detailed in Appendix~\ref{appendix:kz_band}. These effects certainly occur in the experimental data in Fig.~\ref{fig:wide_CDAD}; see also the supplementary spectra in Appendix \ref{appendix:exp_rest}. For example, different photon energies highlight different band dispersions, while the broad distribution of spectral weight for each band is evidence for photoemission from a continuum of states due to the finite $k_z$ resolution. In principle, states with different $k_z$ can have distinct orbital character, which then produces a photon energy dependent CD texture. While we cannot fully exclude this contribution in our experiment, our calculations suggest it is not a dominant factor. For each band in the slab calculation in Fig. \ref{fig:Local_OAM}, the continuum of states corresponding to $k_z$ dispersion have largely the same sign of local OAM. Therefore, simply selecting different $k_z$ in the photomission process is not sufficient to produce the dramatic sign changes observed experimentally.

\begin{figure}
    \centering
        \includegraphics[width=0.5\textwidth]{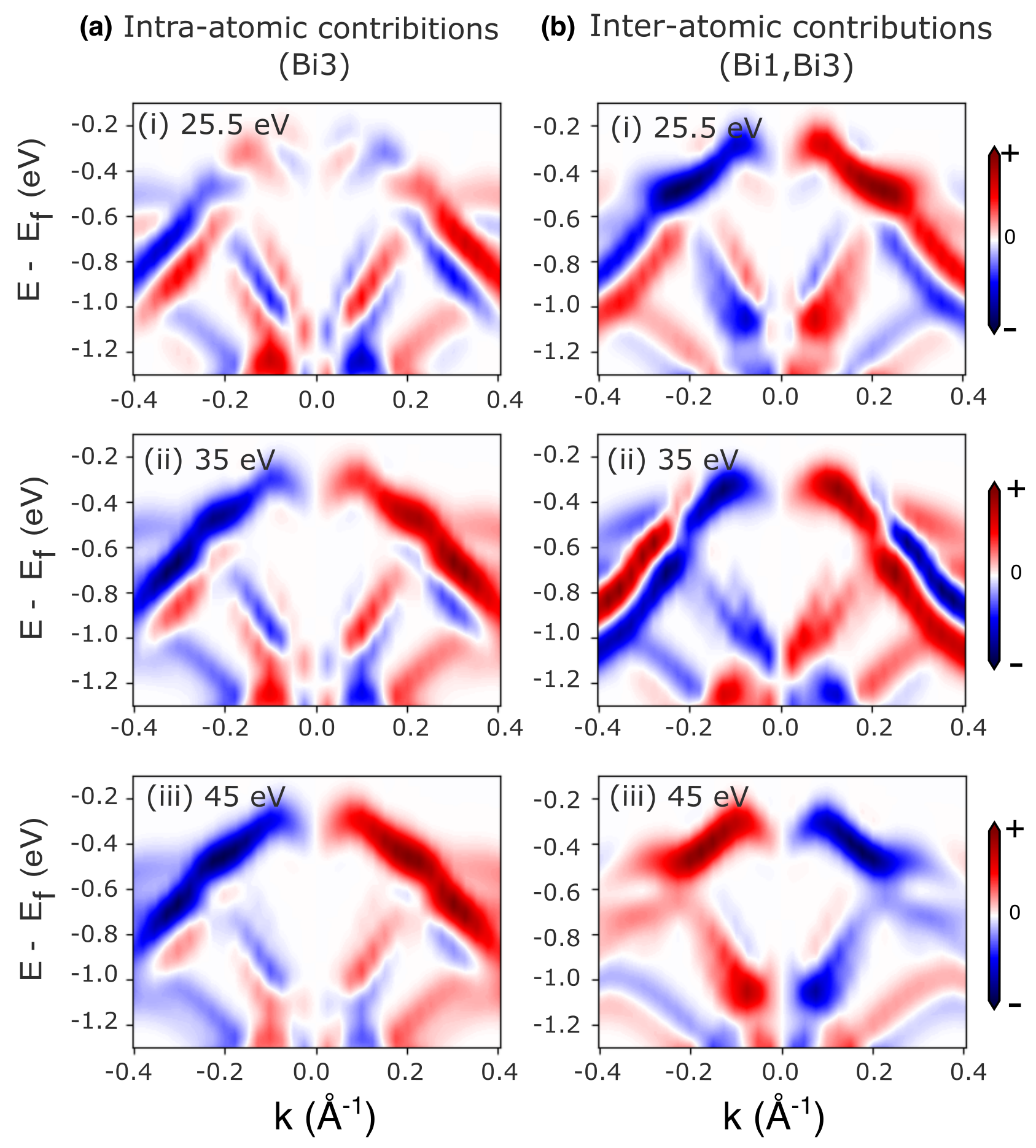}
        \caption{(a) Intra-atomic contribution for Bi3 $I_{\mathrm{intra}}^{\mathrm{Bi3}}$ and (b) inter-atomic interference terms between Bi1 and Bi3 $I_{\mathrm{inter}}^{\mathrm{Bi1,Bi3}}$. Signal for three photon energies are shown to demonstrate the photon energy dependence, over the part of the spectrum highlighted in Fig.~\ref{fig:wide_CDAD}. All the data in this figure is calculated from the bulk bands corresponding to the $k_z$ curve resolved with photon energy $\hbar\omega=$35 eV, with $\vec{k_{\parallel}}$ along $K$-$\Gamma$-$K$ direction. See Appendix~\ref{appendix:kz_band} for detailed discussion.}
    \label{fig:intra_inter}
\end{figure}

On the other hand, the kinetic energy dependent $I_{\mathrm{CD}}(\vec{k}_{\parallel},E)$ in Eq.~\eqref{eq:CDAD_theory_intensity} generally varies with photon energy even within the same set of bands, in agreement with our data. To illustrate this, we examine the intra-atomic and inter-atomic contributions for bulk bands in the limit of $\lambda \rightarrow \infty$. The intra-atomic term $I_{\mathrm{intra}}^{s}(\vec{k}_\parallel,E)$ is directly related to the local OAM through the tensor $T_{smsm'}$ in Eq.~\ref{eq:T_tensor}~\cite{moser_toy_2023,yen_controllable_2024}. Therefore, while the intra-atomic terms reflect the underlying orbital texture, the proportionality factor can be energy dependent. We investigate this computationally for the case of the Bismuth $6p$ orbitals in Fig.~\ref{fig:intra_inter}(a) for three photon energies. Although subtle photon energy dependence is observed, it is not as dramatic as that observed experimentally.


In contrast, the inter-atomic interference terms $I_{\mathrm{inter}}^{s \neq s'}$ vary over the entire photon-energy range due to the phase factor in the tensor $T_{sms'm'}\sim e^{-i\vec{p}\cdot(\vec{r}_s - \vec{r}_{s'})}$ in Eq.~\eqref{eq:T_tensor}. Since photoelectron momentum $\vec{p}$ is a function of $k_\parallel$ and kinetic energy $E$, such interference is strongly momentum- and photon energy-dependent. This is demonstrated in Fig~\ref{fig:intra_inter}(b), where we compute one of the inter-atomic terms $I_{\mathrm{inter}}^{Bi1,Bi3}$ between Bi1 and Bi3 (as labeled in Fig.~\ref{fig:Local_OAM}), showing a strong photon energy dependence over the entire range. 

We conclude that photon energy dependent contributions can originate from three distinct physical mechanisms which compete within our experimental range. However, the $k_z$-dependence and intra-atomic photoemission terms lack the strong photon energy dependence observed experimentally. We therefore conclude that the inter-atomic interference terms are the dominant factor. By carefully selecting photon energies within an appropriate regime (see Appendix~\ref{appendix:relation}) and comparing CDAD at the same $k_z$ slices, it is possible to elucidate the connection between CDAD and the underlying orbital texture with greater clarity~\cite{yen_controllable_2024}.

\subsection{Mechanisms underlying CDAD sign flips in the Dirac topological surface state}

\begin{figure*}
    \centering
        \includegraphics[width=1\textwidth]{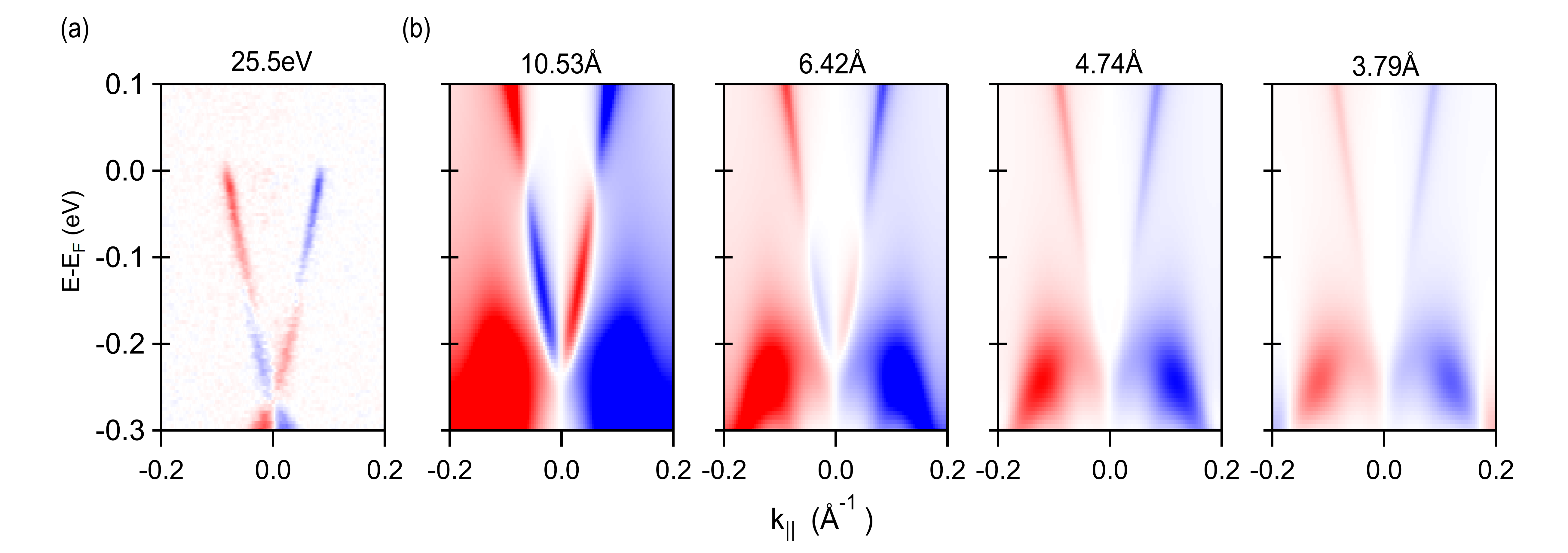}
    \caption{Effect of IMFP on surface state CDAD. (a) Experimental CDAD of the Dirac state along the $K$-$\Gamma$-$K$ direction at 25.5~$eV$. This is compared with (b) the one-step model calculated CDAD extracted for different values of the imaginary part of the time-reversed LEED final state to account for different inelastic mean free paths.}
    \label{fig:IMFP}
\end{figure*}

Following our qualitative understanding of CDAD in the bulk states of $\BiSe$, we turn back to the topological surface state and examine here again the importance of different CDAD contributions. For the Dirac cone a dichroic signal is expected as the surface has broken inversion symmetry, and the Dirac cone has a well-known spin and OAM texture. 
However, it was already noted that final state effects lead to photon energy dependence of the dichroic signal of the surface state \cite{scholz_reversal_2013,vidal_photon_2013}. To capture the details of the photoemission final state beyond the inteference effects discussed above, we perform one-step ARPES calculations based on the KKR framework. The one-step model additionally captures the (spin-dependent) multiple scattering effects of the photoelectrons off the lattice, giving rise to additional phase shifts (see Appendix~\ref{appendix:SPR-KKR} for details). 
%
%
In particular, we focus on the 25.5~eV measurement, which exhibits an unexpected sign-flip within the surface state CDAD, as seen in Fig.~\ref{fig:IMFP}(a) (which is a zoom-in of the left panel in Fig.~\ref{fig:wide_CDAD}) and compare it to one-step calculations. 

First, we examine the effect of the finite IMFP on the surface state CDAD by systematically varying it in Fig.~\ref{fig:IMFP}(b). 
We find a qualitative alteration of the spectral weight and corresponding CDAD texture with IMFP. In particular, the sign-flip of the CD within the Dirac cone is recreated, but only for large enough IMFP. This result exemplifies the importance of the decaying exponential even for wavefunctions that are localized to the surface, as they still have a significant tail going into the bulk.

\begin{figure}
    \centering
        \includegraphics[width=0.5\textwidth]{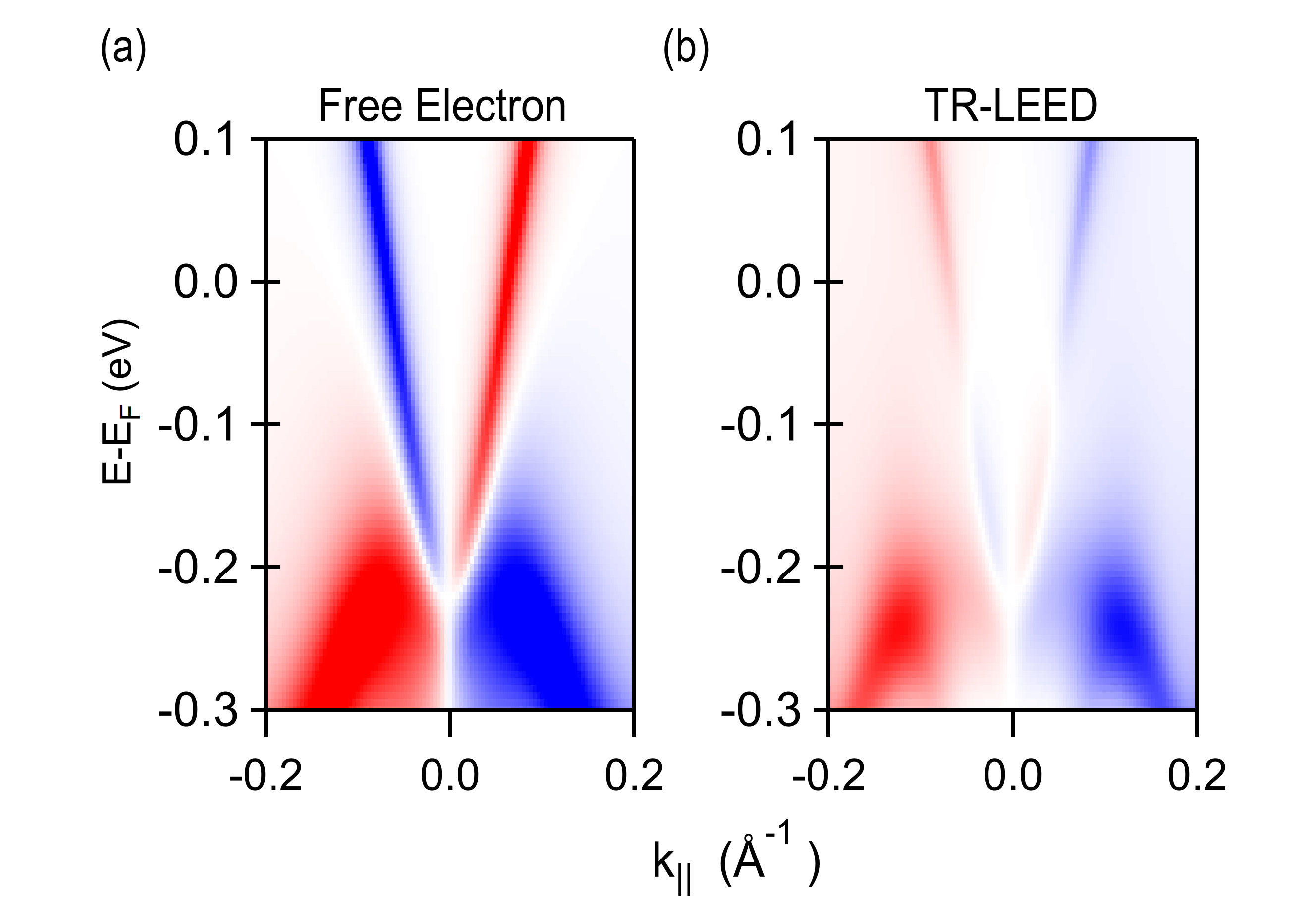}
    \caption{CDAD Comparison of (a) damped free-electron final state (b) TR-LEED final state for the Dirac surface state along the $K$-$\Gamma$-$K$ direction calculated at 25.5~eV.}   
    \label{fig:final_state}
\end{figure}

We can further demonstrate the role of final-state effects by comparing the calculation for a damped free-electron final state to a fully scattered time-reversed LEED (TR-LEED) state.
The TR-LEED final state is composed of a free-electron wave combined with a damped wavefield describing the phase contributions arising from scattering sites inside the solid \cite{Braun2018}. Therefore, TR-LEED states can be used to highlight the contributions of scattering phase shifts. The comparison in Fig.~\ref{fig:final_state} shows that the final state has a strong effect on the overall CDAD, indicating that scattering phase shifts are crucial for its description.
In accordance with previous work~\cite{Schusser2024,schusser_assessing_2022}, it also shows that going beyond the simplified free-electron-like description is necessary to capture ARPES intensity and related observables properly.

\section{Discussion and Conclusions}
We have presented ARPES measurements with circularly polarized light and the resulting CDAD for several photon energies. Despite being subject to both inversion and time-reversal symmetries, the bulk bands exhibit prominent and complex CDAD patterns. The magnitude and sign of the dichroic response vary significantly with momentum and photon energy. The most direct implication of this work is that CD-ARPES cannot be used as a criterion for distinguishing bulk from surface states. 

We presented various sources for the dichroic signal, rooted in the local atomic OAM and manifested in the photoemission process due to its surface sensitivity. Furthermore, we have identified inter-atomic interference as the main contributor to the rich photon energy dependence. We show that the finite IMFP influences even surface-state CDAD, and further identify interference effects stemming from final-state scattering. 

Due to their sensitive dependence on both photon energy and interference/scattering in the photoemission process, these mechanisms are challenging to model quantitatively. Nonetheless, they are unavoidable and must be considered when aiming to extract information on the orbital texture~\cite{boban_scattering_2024} from an ARPES measurement. The interference effects can be reduced by considering additional symmetry operations \cite{beaulieu_revealing_2020}, by choosing photon energies minimizing the CD at high symmetry points \cite{erhardt_bias-free_2024}, or by working in a sufficiently small range of energy and momentum within which interference effects are approximately constant. While CDAD cannot be used to identify polarized states, if the states \emph{are} non-degenerate, the CDAD signal may indeed correspond to the global OAM if interference effects are mitigated. 

While this work shows that CD-ARPES is not a straightforward method to extract OAM, it also reveals the layers of information hidden in the data, which can, in principle, be accessed with further modeling. 
In particular, it gives access to the otherwise "hidden" OAM of the local atomic orbitals, and through interference effects provides access to the photoemission phase which is related to the attosecond-scale delays between different photoemission channels
~\cite{tao_direct_2016}. These subtle delays are the focus of attosecond-science and are linked to, e.g., electronic correlations~\cite{heinrich_attosecond_2021}. However, they have been notoriously hard to measure, in particular in solids~\cite{borrego-varillas_attosecond_2022}. A direct phase-sensitivity via CD-ARPES could provide an alternative route for their quantification.

\section*{Acknowledgments}
The experimental work was supported by the U.S. Department of Energy, Office of Basic Energy Sciences, Division of Materials Science and Engineering under contract DE-AC02-76SF00515. Use of the Stanford Synchrotron Radiation Lightsource, SLAC National Accelerator Laboratory, is supported by the US Department of Energy, Office of Science, Office of Basic Energy Sciences under Contract no. DE-AC02-76SF00515. Y. Y. and M.S. acknowledge support from SNSF Ambizione Grant No. PZ00P2-193527. H. S. acknowledges the support of the Zuckerman STEM Leadership Program, the Young Faculty Award from the National Quantum Science and Technology program of the Israeli Planning and Budgeting Committee, the ERC PhotoTopoCurrent 101078232 and the Israel Science Foundation (grant no. 2117/20).
J.M., S.W.D, J.S and A.P are thankful for the support of the QM4ST project funded by Programme Johannes Amos Commenius, call Excellent Research (Project No. CZ.02.01.01/00/22\_008/0004572).

\appendix

\section{DFT calculations and Wannier tight-binding model for bulk/slab Bi$_2$Se$_3$}
\label{appendix:DFT}
The ground-state band structure of Bi$_2$Se$_3$ was obtained from DFT calculations with the QUANTUM ESPRESSO code~\cite{giannozzi_quantum_2009}. We choose PBE~\cite{perdew_generalized_1996} generalized gradient approximation (GGA) for the exchange-correlation functional including spin-orbit coupling. 

We constructed Wannier functions for the bulk structure using Bi-$6s$, Bi-$6p$, Se-$4s$, and Se-$4p$ with the Wannier90 code~\cite{pizzi_wannier90_2020}. We used the projective Wannier function approach without maximal localization to optimize the match of the Wannier functions with atom-like orbitals with well-defined angular quantum numbers. 

From the bulk Wannier Hamiltonian we constructed a slab supercell of 7 quintuple layers to reproduce the topological surface states. The slab Hamiltonian is then used to calculate the local OAM in Fig.~\ref{fig:Local_OAM}.

\section{Relation between \textit{local} OAM and intra-atomic contribution}
\label{appendix:relation}

    \begin{figure*}
        \centering
        \includegraphics[width=0.9\textwidth]{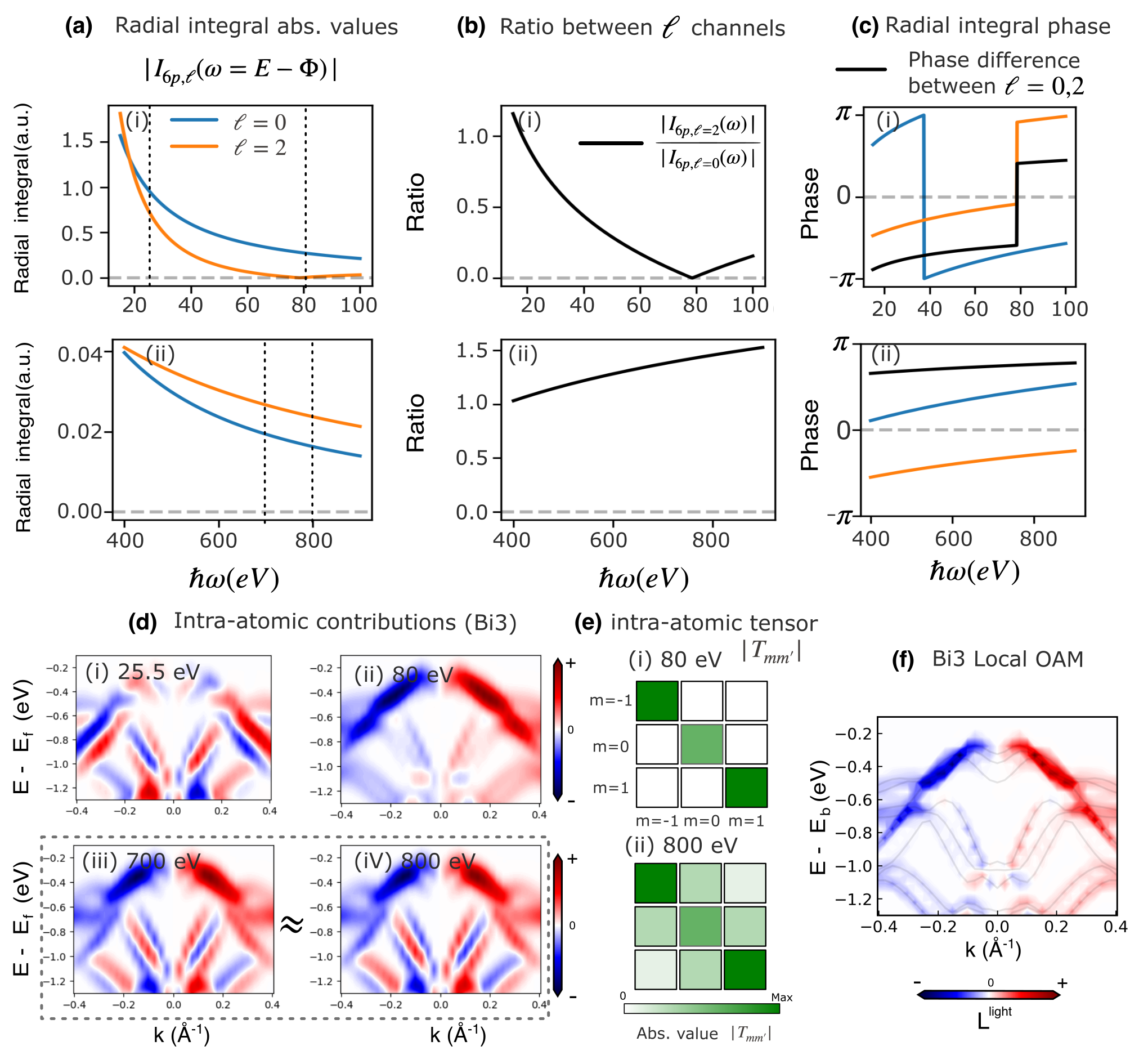}
        \caption{Radial integral $I_{6p,\ell}$ as a function of photon energy of $6p$ orbital solved from atomic all-electron potentials. (a) Absolute values of the radial integral, (b) Ratio between two channels $|I_{6p,\ell=2}|/|I_{6p,\ell=0}|$, and (c) phases of the radial integral for (i) $\hbar\omega=$15 eV to $\hbar\omega=$100 eV and (ii) $\hbar\omega=$400 eV to $\hbar\omega=$900 eV. Here, the final state energy and photon energy are differed by the work function we choose $E=\hbar\omega-\Phi$ with $\Phi=$4.5eV. (d) Bi3 Intra-atomic contributions for bulk bands with photon energy (i) $\hbar\omega=$25.5eV (ii) $\hbar\omega=$80 eV, (iii) $\hbar\omega=$700 eV, and (iV) $\hbar\omega=$800 eV. The corresponding photon energies are labeled with vertical dashed lines in (a). (d) Absolute values of the tensor $|T_{mm'}|=|T_{smsm'}|$ with $s$ denoting Bi3. (f) Local OAM for Bi3 projected onto the light direction.}
    \label{fig:radial_intra}
\end{figure*}
In this appendix, we discuss the details of computing orbital matrix elements and how 
the intra-atomic contributions relate to the \textit{local} OAM. 

Within Wannier-ARPES formalism, photoemission matrix element calculation can be simplified as a summation over orbital matrix elements using atomic center approximation~\cite{yen_first-principle_2024}. For the intra-atomic and inter-atomic contributions in Fig.~\ref{fig:intra_inter}, we apply locally distorted wave approximation to the final states, which takes the effective local potential around each atom into account. Around each atom, the final state wavefunction with momentum $\vec{p}$ is expressed in terms of spherical harmonics as 
    \begin{align}
        \label{eq:final_state}
           \chi_{\vec{p}}(\vec{r}) = 4\pi \sum^\infty_{\ell = 0} \sum^{\ell}_{m=-\ell} i^\ell \frac{w_{\ell}(r,E=\vec{p}^2/2)}{r} Y_{\ell m}(\Omega_{\vec{r}}) Y^*_{\ell m}(\Omega_{\vec{p}}) \ .
    \end{align}
    The radial wave function $w_{\ell}(r,E)$ is the scattering state with kinetic energy $E=\vec{p}^2/2$, calculated from the radial Schr{\"o}dinger equation with the all-electron Kohn-Sham potential. In Eq.~\eqref{eq:final_state}, $\Omega_{\vec{r}}$ and $\Omega_{\vec{p}}$ refer to the solid angle of the vectors $\vec{r}$ and $\vec{p}$. The orbital matrix elements in Eq.~\ref{eq:mel_dip_aca_orb} are then calculated as  
    \begin{align}
        \label{eq:dist_wave}
            M^{\mathrm{orb}}_j(\vec{k}_{\parallel},E) 
             =4\pi & \sum_{\ell m} (-1)^{m} (-i)^\ell Y_{\ell m} (\Omega_{\vec{p}})\nonumber \\ &\times C_{\ell m, \ell_j m_j}(\vec{\epsilon}) I_{j,\ell}(E) \ .
    \end{align}
    The Wannier functions in Eq.~\eqref{eq:mel_dip_aca_orb} are approximated as $ \phi_j(\vec{r})=u_{j}(r)Y_{\ell_j,m_j}(\Omega_{\vec{r}})/r$, with the radial function $u_{j}(r)$ computed from radial Schr{\"o}dinger equation with isolated atoms. The summation runs over final state channels $\ell$ and $m$. The dipole selection rules $\ell_j \rightarrow \ell=\ell_{j}\pm1$ and  $m_j \rightarrow m=m_j,m_{j}\pm1$ are reflected in the angular coefficient $C_{\ell m, \ell_j m_j}(\vec{\epsilon})$, which is computed from the light polarization $\vec{\epsilon}$ and the Clebsch-Gordan coefficients. The radial integral $I_{j,\ell}(E)$ is defined as
    \begin{align}
        \label{eq:radial}
        I_{j,\ell}(E)= \int^\infty_0 dr\, w^*_{\ell}(r,E) r u_j(r) \ . 
    \end{align}
    The kinetic energy $E=\hbar\omega - \Phi$ is counted from Fermi level and related to photon energy by the work function $\Phi$=4.5 eV. 

    As we discussed in Section~\ref{s:wannier_arpes}, the detailed relation between local OAM and the intra-atomic contribution is intricate, and it depends on the experimental geometry, orbital texture~\cite{yen_controllable_2024,moser_toy_2023}, and even the photon energy. Here, we discuss the photon energy dependence of the intra-atomic OAM contributions, using Bismuth 6p orbitals as an illustration. 
     
    In Fig.~\ref{fig:radial_intra}(a)-(c), we show the computed radial integral for Bismuth 6p orbitals using atomic-all electron potential. Two photon energy ranges (i) $\hbar\omega=$ 15 to 100 eV and (ii) $\hbar\omega=$ 400 to 900 eV are displayed to present the contrast between our experimental low photon energies and the large photon energy regime. Due to the dipole selection rules, the only available final-state channels are the ones with angular momentum quantum numbers $\ell=0,2$. The absolute values of the two channels decrease rapidly in our experimental range ($\hbar\omega=$ 25.5 - 50 eV). In Fig.~\ref{fig:radial_intra}(a), $\ell=2$ channel falls to zero at $\hbar\omega=$80 eV, corresponding to the so-called Cooper minimum~\cite{cooper1962photoionization}. With large photon energies, the absolute value of the $\ell=2$ channel exceeds the $\ell=0$ channel but the ratio between them becomes flattened and insensitive to photon energy, as shown in Fig.~\ref{fig:radial_intra}(b). In Fig.~\ref{fig:radial_intra}(c), the phase of $\ell=2$ channel jumps at the Cooper minimum, and the difference between the two phases in high photon energies remains stable. 

    In the experimental range ($\hbar\omega=$ 25.5 - 50 eV), the ratio between two channels varies rapidly. Thus, the resulting intra-atomic contributions have a sensitive dependence on photon energy. On the other hand, in the large photon energy limit, the stable ratio and phase difference between channels lead to nearly photon energy independent intra-atomic contributions. We demonstrate these dependencies for the Bi3 in the bulk limit ($\lambda \rightarrow \infty$) in Fig.~\ref{fig:radial_intra}(d). The intra-atomic terms within the experimental range show photon energy dependence, as demonstrated in Fig.~\ref{fig:intra_inter}(a). Here we show again the $\hbar\omega=$25.5 eV result, which has a very different texture from all the other photon energies in  Fig.~\ref{fig:intra_inter}(a) and Fig.~\ref{fig:radial_intra}(d). At the Cooper minima $\hbar\omega=$80 eV, $\ell=2$ channel does not contribute due to vanishing radial integral, leaving the only present process to be $\ell_j=1$ (p) $\rightarrow$ $\ell=0$ (s). Effectively, the tensor $T_{mm'}=T_{smsm'}$ with $s=$Bi3 as defined in Eq.\eqref{eq:T_tensor} becomes diagonal~\cite{moser_toy_2023} and proportional to the atomic OAM tensor in Eq.~\eqref{eq:local_oam} along the quantization axis defined to be the incoming light direction. The tensor is demonstrated with a schematic in Fig.~\ref{fig:radial_intra}(e)-(i). As a result, the intra-atomic contribution shows \textit{direct proportionality} to the local OAM in Fig.~\ref{fig:radial_intra}(f). 

    At large photon energies, the intra-atomic contributions show photon energy independent textures in Fig.~\ref{fig:radial_intra}(d)-(iii) and (d)-(iv). The two final-state channels remain finite with a stable ratio, leading to some finite off-diagonal terms in the tensor $T_{mm'}$ demonstrated in Fig.~\ref{fig:radial_intra}(e)-(ii). Therefore, the intra-atomic contributions approximately reflect the local OAM in Fig.~\ref{fig:radial_intra}(f) but lose the direct proportionality.

\section{Layer-resolved OAM and surface symmetry breaking}
\label{appendix:layer_resolved}

 \begin{figure}
        \includegraphics[width=0.5\textwidth]{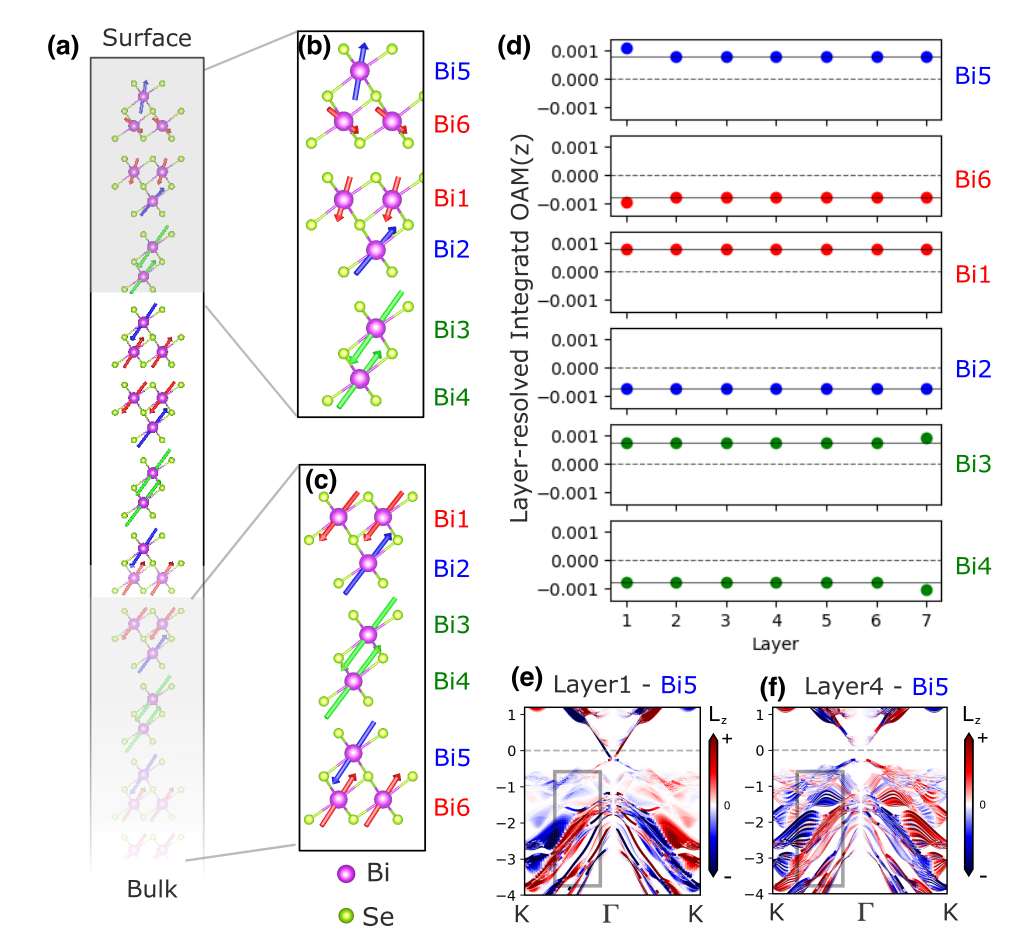}
    \caption{Layer-resolved OAM for a 7-layer slab. (a) Local OAM vectors for the whole slab. (b)-(c) Zoom-in OAM vectors for surface and bulk. (d) Layer-resolved $z$ component OAM for each Bismuth atom in a unit cell. (e)-(f) Local OAM for Bi5 in the surface layer (layer 1) and bulk (layer 4), plotted on top of slab bands. The OAM vectors in (a)-(c) are integrated values over the black box range in (e)-(f). }
    \label{fig:layer-resolved}
\end{figure}
The surface breaks the inversion symmetry inherently and can modify the surface electronic structure. In this appendix, we show the effects of such surface symmetry breaking on local OAM for the atoms close to the surface. Now we show again the calculated local OAM in the 7-layer slab in Fig.~\ref{fig:layer-resolved}(a), with a focus on the surface layer (Fig.~\ref{fig:layer-resolved}(b)). The OAM in the surface layer breaks the perfect symmetry of the bulk OAM  (Fig.~\ref{fig:layer-resolved}(c)). In Fig.~\ref{fig:layer-resolved}(d), the $z$ component of local OAM vectors for every atom in the 7 layers slab is shown, where we can again see the local OAM for atoms closer to the surface differing from the bulk value (Bi5/Bi6 for the top surface and Bi3/Bi4 for the bottom surface). From the slab-band resolved local OAM, there is also a clear difference between the surface and the bulk in Fig.~\ref{fig:layer-resolved}(e) and (f). 

In section~\ref{s:finite_cd}, we argue that inelastic mean free path profile leads to the breaking of pairwise cancellation and non-vanishing CD in the bulk bands. Although the surface symmetry breaking depicted in Fig.~\ref{fig:layer-resolved} also breaks the pairwise cancellation, we stress that this is only a secondary effect, as such a deviation of the surface local OAM is small and occurs only for the top few atoms.

\section{Photon energy dependent $k_z$ resolution}
  \label{appendix:kz_band}

  \begin{figure}

    \includegraphics[width=0.5\textwidth]{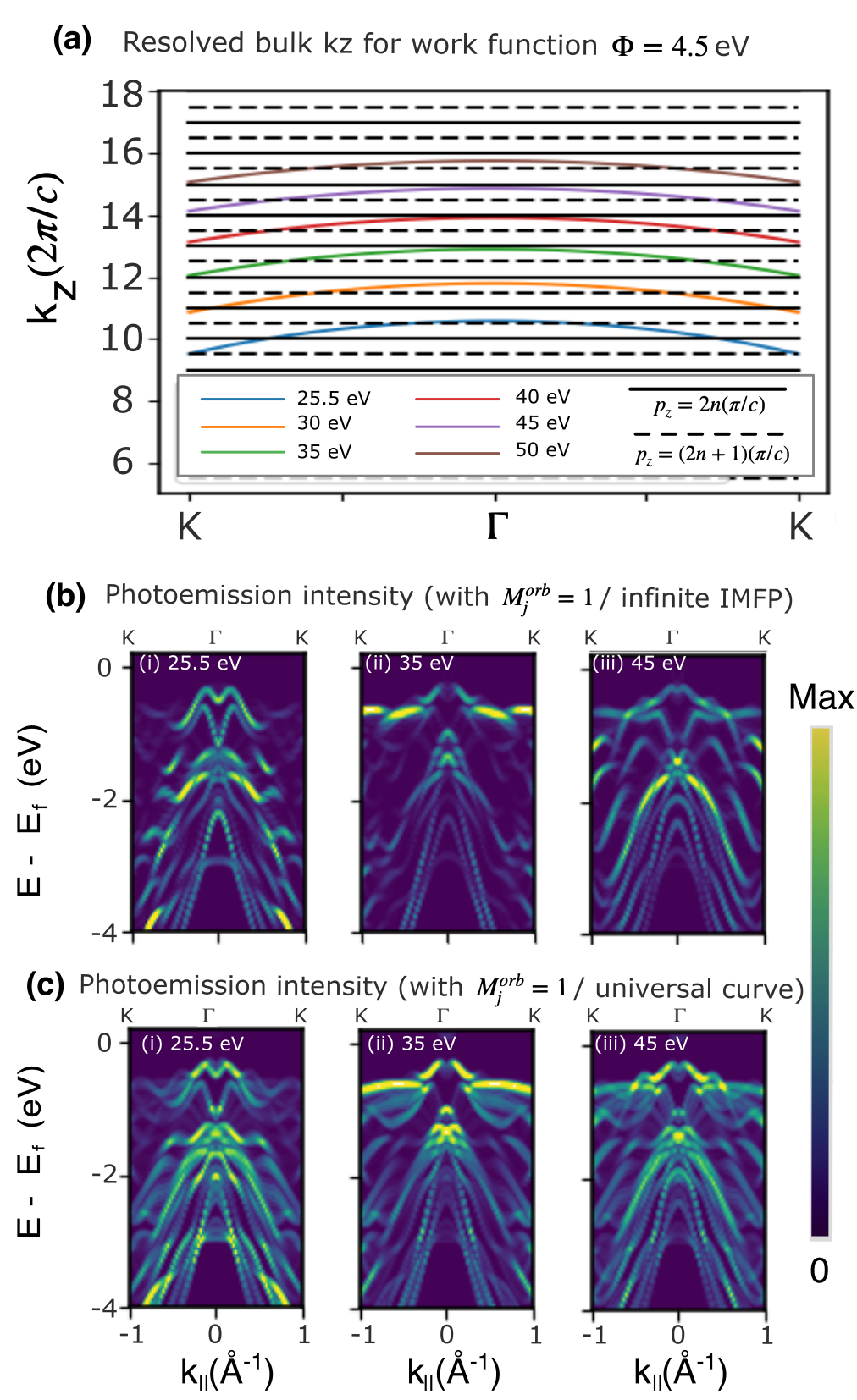}
    \caption{(a) Resolved $k_z$ calculated with different photon energy with work function $\Phi=4.5$eV. The solid lines correspond to $k_z=2n\frac{\pi}{c}$ with integer $n$ and out-of-plane lattice constant $c$, while the dashed line denotes the zone boundary $k_z=(2n+1)\frac{\pi}{c}$. (b) Calculated photoemission intensity from a 10 layers slab with $M_j^{orb}=1$ and in the bulk limit with infinite IMFP. (c) Calculated photoemission intensity with IMFP from the universal curve.}
    \label{fig:kz_band}
\end{figure}
  The out-of-plane momentum $p_z$ of the photoelectron in vacuum is related to the probed crystal momentum $k_z$ in the solid via the inner potential $V_0$: $p^2_z /2 = k^2_z/2-V_0$. Note that $\vec{p}$ enters into the matrix element phase factor $e^{-i\vec{p}\cdot \vec{r}_j}$, while $\vec{k}$ is more directly associated with the dispersion of electronic states. Both depend on photon energy $\hbar\omega$, in-plane momentum $\vec{k_\parallel}$, and band energy $\epsilon_\alpha(\vec{k}_{\parallel})$ via:
\begin{equation}
    \label{eq:inner_potential}
        \begin{aligned}
        E & =p_z^2/2 + \vec{k}_{\parallel}^2/2 \\ & = k_z^2/2 + \vec{k}_{\parallel}^2 - V_0 \\ & =\hbar\omega-\Phi+\epsilon_\alpha(\vec{k}_{\parallel}) \ ,
        \end{aligned}
\end{equation}
where $E$ is the kinetic energy, and $\Phi$ is the work function. $V_0$ is typically determined experimentally through the periodicity of a $\hbar\omega$-dependent dataset \cite{damascelli_probing_2004}. For demonstration purposes, the calculations shown here take $V_0 = 0$ and $\Phi=4.5$ eV. The band energy $\epsilon_\alpha(\vec{k}_{\parallel})$ is defined relative to the Fermi energy.
    
    In Fig.~\ref{fig:kz_band}(a), we illustrate such $k_z$-dependence along K-$\Gamma$-K with fixed $\epsilon_\alpha(\vec{k}_{\parallel})=-1$ eV, which shows that bulk bands at different $k_z$ are highlighted with different $\hbar\omega$. To show the effect on the spectra, we compute photoemission intensity with Wannier-ARPES formulation for a 10-layer slab. The effects of matrix elements are excluded by replacing $M_j^{orb}=1$ in Eq.\eqref{eq:mel_dip_aca}, such that the intensity modulation is purely due to the phase factor. In the limit ($\lambda \rightarrow \infty$), shown in Fig.~\ref{fig:kz_band}(b), the intensity is distributed among slab bands which correspond to bulk bands with the effective $k_z$  from panel (a). If a realistic IMFP is included with values from the universal curve, the attenuation factor $e^{z_j/\lambda}$ broadens the $k_z$ resolution as shown in Fig.~\ref{fig:kz_band}(c). Therefore, a continuum of bulk bands at different $k_z$ is probed in the ARPES spectrum~\cite{day_looking_2023}.


\section{Photoemission with different polarizations}
\label{appendix:exp_rest}
\begin{figure*}

    \includegraphics[width=1\textwidth]{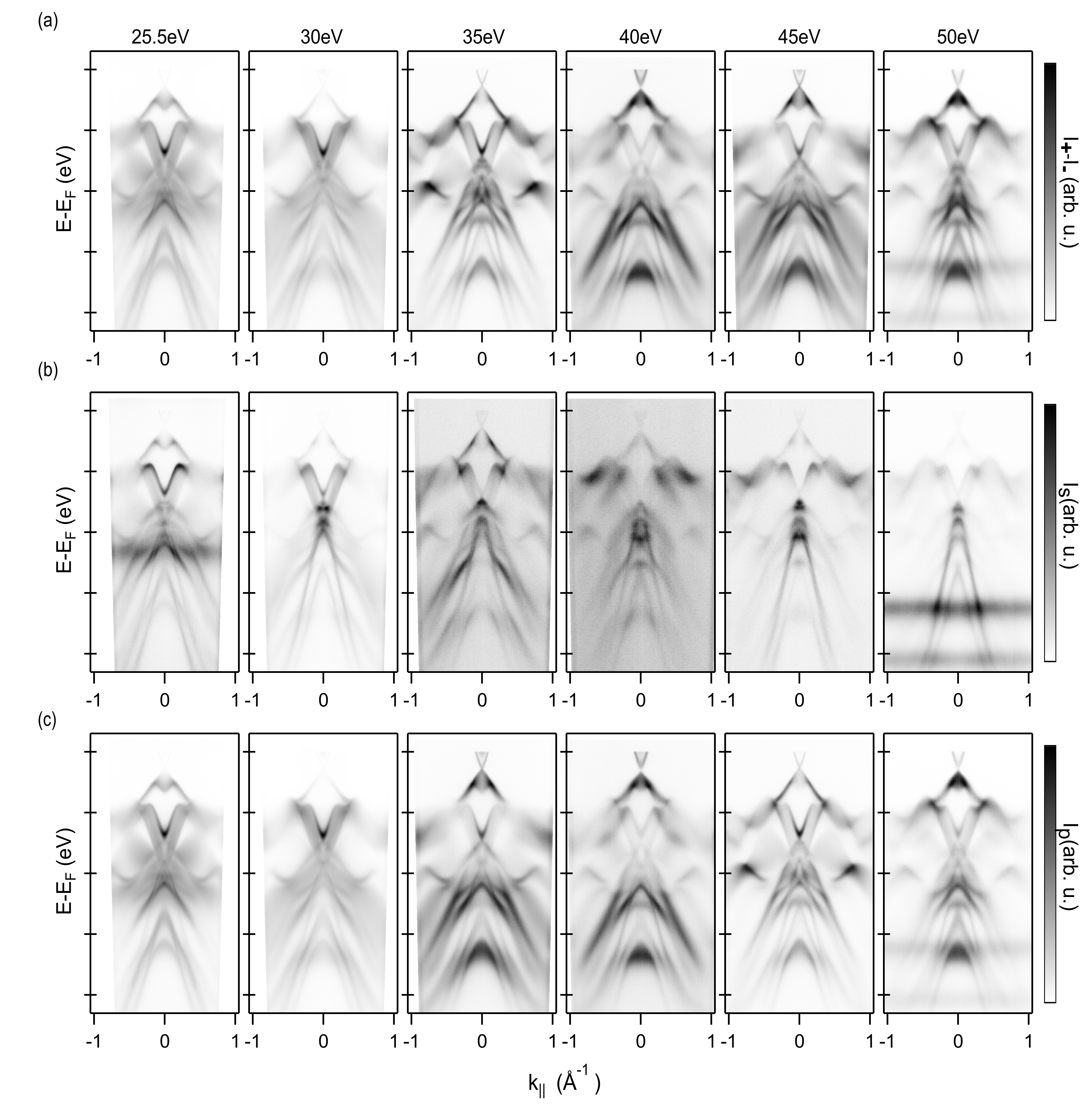}
    \caption{ARPES spectra along the $\Gamma$-K direction. (a) sum of the two circular polarizations. (b) S polarization (parallel to the sample's surface in Fig.\ref{fig:Exp_Geo}a). (c) P polarization (parallel to the blue mirror plane in Fig.\ref{fig:Exp_Geo}a)}
    \label{fig:All_ARPES}
\end{figure*}
Experimental ARPES spectra along the $\Gamma$-K direction, obtained for different light polarizations, are presented in Fig.\ref{fig:All_ARPES}. The summed spectra of the two circularly polarized  Fig.\ref{fig:All_ARPES}(a) have a symmetric structure around the $\Gamma$ point, similarly to the vertical (S) and horizontal (P) linear polarization spectra (Fig.\ref{fig:All_ARPES}(b) and (c), respectively). Spectra for all polarizations exhibit significant photon energy dependence, with different bands highlighted at each one. All spectra are normalized by the photon flux as mentioned in Section \ref{s:exp}.

\section{SPR-KKR calculations}
\label{appendix:SPR-KKR}
We employed the Spin-Polarized Relativistic Korringa-Kohn-Rostoker (SPR-KKR) calculations to simulate the ARPES spectra. The SPR-KKR package is based on Green's function formalism and allows for an accurate description of the photoemission matrix element by means of a one-step model \cite{EbertKodderitzschMinar2011,MinarBraunMankovskyEbert2011,Braun2018}. This fully relativistic simulation includes all surface and interference phenomena, such as crystal termination, bulk sensitivity, and phase shifts due to the scattering of the final state. To simulate the semiinfinite Bi$_2$Se$_3$, we employed local density approximation (LDA) as the exchange-correlation potential. We used the angular-momentum cutoff $l_\mathrm{max}$= 3 to account for the allowed transitions imposed by the dipole selection rules. Our model based on atomic spheres approximation converged with 43 reciprocal lattice vectors, the initial state broadening was implemented via the imaginary part of the potential V$_i$=0.05~$eV$ and to tune the final state damping and the corresponding IMFP, we varied the imaginary part of the potential of the final state.

\clearpage

\bibliography{cdad_refs_formatted}
\selectlanguage{american}%
\end{document}